\documentclass[twocolumn]{aastex61}
\usepackage{amsmath,amstext}
\usepackage[T1]{fontenc}
\usepackage{apjfonts} 
\hypersetup{citecolor=blue,linkcolor=red}
\usepackage[all]{hypcap}

\def\cfa{Harvard-Smithsonian Center for Astrophysics, 60 Garden Street, Cambridge, MA 02138, USA}

\def\gw{GW170817}
\def\ngc{NGC\,4993}

\shorttitle{The Host Galaxy of \gw}
\shortauthors{Blanchard et al.}

\begin{document}

\title{The Electromagnetic Counterpart of the Binary Neutron Star Merger LIGO/VIRGO GW170817. VII. Properties of the Host Galaxy and Constraints on the Merger Timescale}

\author{P. K. Blanchard}
\affiliation{\cfa}
\affiliation{NSF GRFP Fellow}
\email{pblanchard@cfa.harvard.edu}
\author{E. Berger}
\affiliation{\cfa}
\author{W. Fong}
\affiliation{Hubble Fellow}
\affiliation{Center for Interdisciplinary Exploration and Research in Astrophysics (CIERA) and Department of Physics and Astronomy, Northwestern University, Evanston, IL 60208, USA}
\author{M. Nicholl}
\affiliation{\cfa}
\author{J. Leja}
\affiliation{\cfa}
\author{C. Conroy}
\affiliation{\cfa}
\author{K. D. Alexander}
\affiliation{\cfa}
\author{R. Margutti}
\affiliation{Center for Interdisciplinary Exploration and Research in Astrophysics (CIERA) and Department of Physics and Astronomy, Northwestern University, Evanston, IL 60208, USA}
\author{P. K. G. Williams}
\affiliation{\cfa}
\author{Z. Doctor}
\affiliation{Department of Physics, University of Chicago, Chicago, IL 60637, USA}
\author{R. Chornock}
\affiliation{Astrophysical Institute, Department of Physics and Astronomy, 251B Clippinger Lab, Ohio University, Athens, OH 45701, USA}
\author{V. A. Villar}
\affiliation{\cfa}
\author{P. S. Cowperthwaite}
\affiliation{\cfa}
\author{J. Annis}
\affiliation{Fermi National Accelerator Laboratory, P. O. Box 500, Batavia, IL 60510, USA}
\author{D. Brout}
\affiliation{Department of Physics and Astronomy, University of Pennsylvania, Philadelphia, PA 19104, USA}
\author{D. A. Brown}
\affiliation{Department of Physics, Syracuse University, Syracuse, NY 13224, USA}
\author{H.-Y. Chen}
\affiliation{Department of Astronomy and Astrophysics, University of Chicago, Chicago, IL 60637, USA}
\author{T. Eftekhari}
\affiliation{\cfa}
\author{J. A. Frieman}
\affiliation{Fermi National Accelerator Laboratory, P. O. Box 500, Batavia, IL 60510, USA}
\affiliation{Kavli Institute for Cosmological Physics, The University of Chicago, Chicago, IL 60637, USA}
\author{D. E. Holz}
\affiliation{Enrico Fermi Institute, Department of Physics, Department of Astronomy and Astrophysics,\\and Kavli Institute for Cosmological Physics, University of Chicago, Chicago, IL 60637, USA}
\author{B. D. Metzger}
\affiliation{Department of Physics and Columbia Astrophysics Laboratory, Columbia University, New York, NY 10027, USA}
\author{A. Rest}
\affiliation{Space Telescope Science Institute, 3700 San Martin Drive, Baltimore, MD 21218, USA} 
\affiliation{Department of Physics and Astronomy, The Johns Hopkins University, 3400 North Charles Street, Baltimore, MD 21218, USA}
\author{M. Sako}
\affiliation{Department of Physics and Astronomy, University of Pennsylvania, Philadelphia, PA 19104, USA}
\author{M. Soares-Santos}
\affiliation{Department of Physics, Brandeis University, Waltham, MA 02454, USA}
\affiliation{Fermi National Accelerator Laboratory, P. O. Box 500, Batavia, IL 60510, USA}

\begin{abstract}

We present the properties of \ngc, the host galaxy of \gw, the first gravitational wave (GW) event from the merger of a binary neutron star (BNS) system and the first with an electromagnetic (EM) counterpart. We use both archival photometry and new optical/near-IR imaging and spectroscopy, together with stellar population synthesis models to infer the global properties of the host galaxy.  We infer a star formation history peaked at $\gtrsim 10$ Gyr ago, with subsequent exponential decline leading to a low current star formation rate of 0.01 M$_{\Sun}$ yr$^{-1}$, which we convert into a binary merger timescale probability distribution.  We find a median merger timescale of $11.2^{+0.7}_{-1.4}$ Gyr, with a 90\% confidence range of $6.8-13.6$ Gyr.  This in turn indicates an initial binary separation of $\approx 4.5$ R$_{\Sun}$, comparable to the inferred values for Galactic BNS systems.  We also use new and archival {\it Hubble Space Telescope} images to measure a projected offset of the optical counterpart of $2.1$ kpc (0.64$r_{e}$) from the center of \ngc\ and to place a limit of $M_{r} \gtrsim -7.2$ mag on any pre-existing emission, which rules out the brighter half of the globular cluster luminosity function.  Finally, the age and offset of the system indicates it experienced a modest natal kick with an upper limit of $\sim 200$ km s$^{-1}$.  Future GW$-$EM observations of BNS mergers will enable measurement of their population delay time distribution, which will directly inform their viability as the dominant source of $r$-process enrichment in the Universe.   

\end{abstract}


\section{Introduction}
 
The recent discovery of gravitational waves (GWs) from binary black hole (BBH) mergers \citep{ligo2,ligo1,ligo3} has launched a new era of astronomy.  However, realizing the full potential of GW astronomy for advancing our knowledge of the formation of compact object binaries requires the observation of electromagnetic (EM) counterparts and hence precise positions and association with specific galaxies and stellar populations. While BBH mergers are not expected to produce EM signals, a wide range of EM counterparts are expected for binary systems containing at least one neutron star \citep{MB2012}.  

On 2017 August 17 at 12:41:04 UT the Advanced Laser Interferometer Gravitational-Wave Observatory and Advanced Virgo interferometer (ALAV) discovered the first GW event from the inspiral and merger of two neutron stars \citep[\gw;][]{ALVgcn,ALVDetection}.  A short burst of gamma-rays (GRB\,170817) was independently discovered from the same sky location with a delay of about 2 s by \textit{Fermi}-GBM \citep{GBMdetection} and \textit{INTEGRAL} \citep{INTEGRALdetection}.  About 0.5 days after the GW trigger our group used the Dark Energy Camera on the Blanco 4 m telescope to discover an optical counterpart ($i \approx 17.48$ and $z \approx 17.59$ mag) associated with the galaxy NGC\,4993 at a distance of $d\approx 39.5$ Mpc \citep{DECAMgcn,DECamPaper1}, which was independently discovered by \citet[][dubbed SSS17a]{SWOPEpaper,SWOPEgcn} and \citet[][dubbed DLT17ck]{DLT40gcn}.  The transient is also known as AT\,2017gfo.  The companion papers in this series present strong evidence that the optical counterpart is due to kilonova emission \citep{DECamPaper4,DECamPaper2,DECamPaper3}, with little or no contribution from a GRB afterglow due to viewing angle effects \citep{DECamPaper6,DECamPaper5}.       

Until now, the only observational data informing the formation and evolution of binary neutron star (BNS) systems has been through studies of the Galactic population of BNS systems \citep[e.g.][]{SB1976,Burgay2003,Kalogera2004,Kalogera2007,KS2008} and short GRBs \citep[][and references therein]{Berger2014}.  Numerous open questions remain related to the initial conditions, rate, and population properties of BNS systems, as well as their eventual mergers and role in galactic $r$-process enrichment.  For example, the distribution of delay times (i.e., the sum of the evolutionary time to form a BNS system and its time to merge) is a key output of population synthesis simulations \citep[e.g.][]{VT2003,Belczynski2006,Dominik2012}.  Similarly, the observed locations of short GRBs within their hosts provides constraints on natal kicks and the possibility of globular clusters as formation sites \citep{Fong2010,Church2011,FB2013}.  

Here, we use our follow-up observations and archival data of \ngc\ to measure the precise location of the BNS system at the time of merger and to infer the physical properties of the host, in particular its star formation history, which serves as a proxy for the BNS merger delay time, and hence the initial binary separation.  We compare these results to Galactic BNS systems and results from population synthesis models.

Throughout the Letter, we use AB magnitudes corrected for Galactic extinction, with $E(B-V) = 0.105$ \citep{SF2011}, and the following cosmological parameters: $H_{0} = 67.7$ km s$^{-1}$ Mpc$^{-1}$, $\Omega_{m} = 0.307$, and $\Omega_{\Lambda} = 0.691$ \citep{Planck2016}. 

\section{Observations and Archival Data}

\subsection{\textit{Hubble Space Telescope (HST)} Observations}
\label{sec:hst}

As described in \citet{DECamPaper2} we obtained \textit{HST} Target-of-Opportunity observations of the optical counterpart of \gw\ on 2017 August 28 using the Advanced Camera for Surveys (ACS) with the F475W, F625W, F775W, and F850LP filters, the Wide Field Camera 3 (WFC3) IR channel with the F160W and F110W filters, and the WFC3 UVIS channel with the F336W filter (PID: 15329; PI: Berger).  The data analysis is described in \cite{DECamPaper2}.  

\begin{figure*}[ht!]
\begin{center}
\includegraphics[scale=0.19]{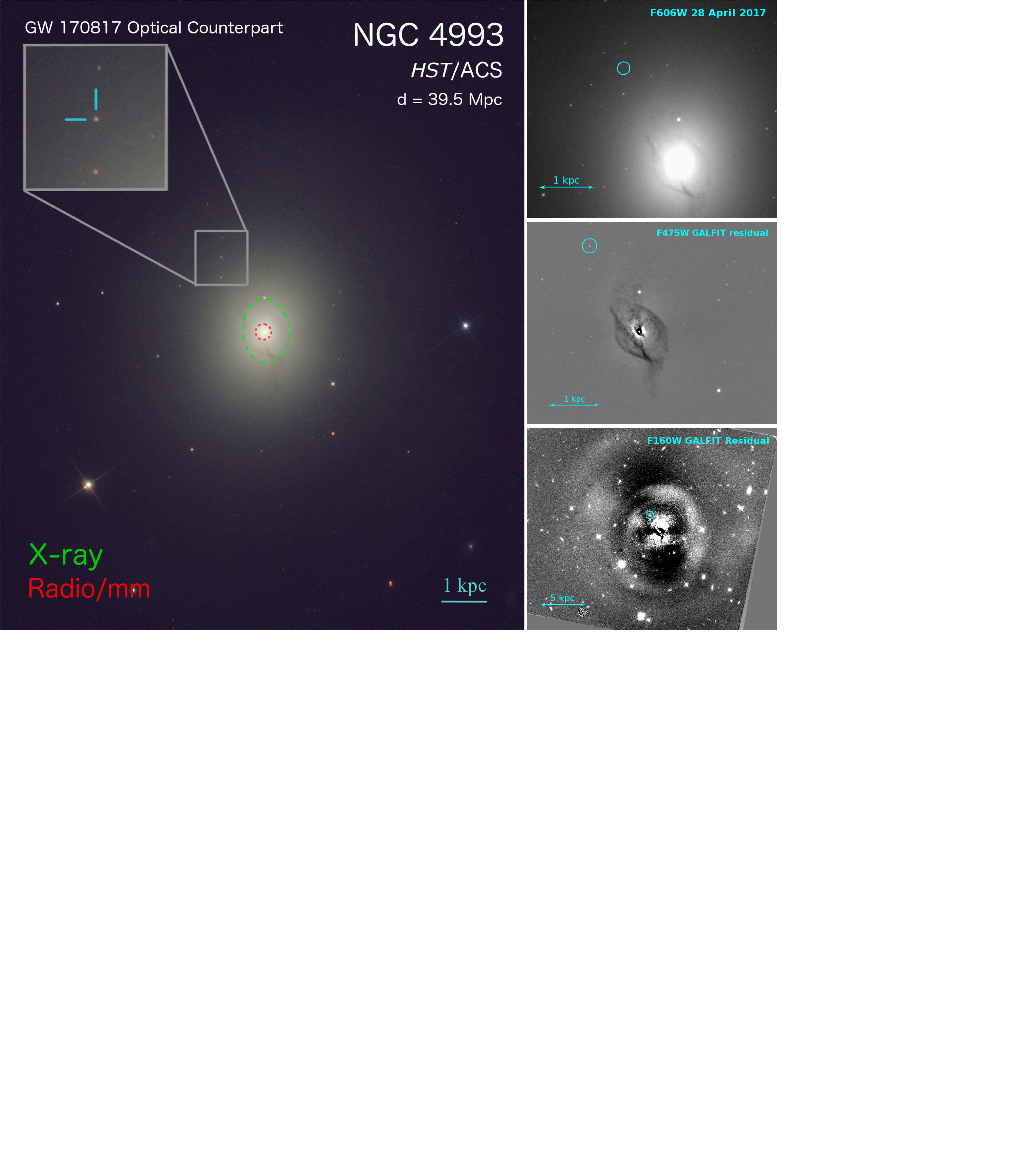}
\end{center}
\caption{\textit{Left}: Color image of \ngc\ created from filtered \textit{HST}/ACS images (F850LP, F625W, F475W).  The inset shows the optical counterpart of \gw\ and the dashed green ellipse (90\% confidence region) and dashed red circle (10$\sigma$ radius for clarity) mark the locations of the X-ray \citep{DECamPaper5} and  millimeter and radio sources \citep{DECamPaper6}, respectively, associated with the host galaxy.  \textit{Top Right}: Archival \textit{HST}/ACS image of \ngc\ from 2017 April 28 exhibits no underlying point source at the position of the optical counterpart (circle) to a limit of $M_{\rm F606W} = -7.2$ mag.  \textit{Middle Right}: GALFIT residual image in the ACS/F475W filter showing the dust structure surrounding the nucleus. \textit{Bottom Right}: GALFIT residual image in the WFC3/F160W filter showing the presence of concentric shells and azimuthal variations.  Dust and shell structure may be indicative of past galaxy mergers.  All images are aligned with North up and East to the left.}
\label{fig:hst}
\end{figure*}

In Figure~\ref{fig:hst}, we show a color image of \ngc\ with an inset showing the location of the optical counterpart of \gw\ ($m_{\rm F625W} \approx 22.9$ mag at this epoch) created using our 2017 August 28 \textit{HST}/ACS images (F850LP, F625W, and F475W).  The galaxy exhibits a smooth surface brightness profile typical of elliptical galaxies, but with a complex dust structure near the nucleus.  

We also retrieved and analyzed an archival observation of \ngc\ from 2017 April 28 using ACS/WFC with the F606W filter (PID: 14840; PI: Bellini), which allows for an assessment of an underlying source at the location of the optical counterpart.  We determine the exact location of the optical counterpart in the archival image by performing astrometry relative to our {\it HST} images.  The resulting astrometric uncertainty is only $0.0075''$ ($1\sigma$) corresponding to $0.2$ pixels.  No obvious source is seen at the location of the optical counterpart (Figure \ref{fig:hst}); the region is  dominated by the background galaxy light.  To obtain a limit on the presence of a point source we use the IRAF/{\tt psf} task to create a point-spread function from the image, and the IRAF/{\tt addstar} task to then  inject fake point sources of varying brightness at the optical counterpart's location.  We find a $5\sigma$ upper limit of $m_{\rm F606W}\gtrsim 26.0$ mag for a point source, corresponding to $M_{\rm F606W}\gtrsim -7.2$ mag at the distance of \ngc.

\subsection{Additional Archival Data}
\label{archival}

For the purpose of modeling the host galaxy spectral energy distribution (SED), we retrieved archival observations of NGC\,4993, including UV and IR photometry from the \textit{GALEX}, 2MASS, and \textit{WISE} catalogs via the NASA/IPAC Extragalactic Database.  For optical data we used deep $grizy$ stacks from the Pan-STARRS1 3$\pi$ survey \citep{Chambers2016,Waters2016} and performed photometry using {\tt SExtractor} \citep{SExtractor}.  We use the {\tt MAG\_AUTO} magnitudes, which are measured using Kron apertures.  The photometry is summarized in Table~\ref{tab:phot}.  

\capstartfalse
\begin{deluxetable}{ccc}
\tablecolumns{3}
\tabcolsep0.1in\footnotesize
\tablewidth{0pc}
\tablecaption{X-Ray to Radio Photometry of \ngc\    
\label{tab:phot}}
\tablehead {
\colhead {Instrument}   &
\colhead {Band}     &
\colhead {Magnitude/Flux}             
}   
\startdata
\textit{Chandra} & $2-10$ keV & $1.7 \times 10^{-14}$ \\
\textit{GALEX} & FUV & $>18.86$ \\
\textit{GALEX} & NUV & 17.82 (0.09) \\
PS1 & $g$ & 12.80 (0.02) \\
PS1 & $r$ & 12.16 (0.01) \\
PS1 & $i$ & 11.81 (0.01) \\
PS1 & $z$ & 11.57 (0.01) \\
PS1 & $y$ & 11.36 (0.02) \\
2MASS & J & 10.98 (0.02) \\
2MASS & H & 10.82 (0.02) \\
2MASS & K & 11.02 (0.02) \\
\textit{WISE} & W1 & 11.92 (0.01) \\
\textit{WISE} & W2 & 12.59 (0.01) \\
\textit{WISE} & W3 & 13.70 (0.04) \\
\textit{WISE} & W4 & 13.86 (0.18) \\
ALMA & 97.5 GHz & 210 (20) \\
VLA & 15.0 GHz & 295 (18) \\
VLA & 10.0 GHz & 288 (20) \\
VLA & 9.7 GHz & 250 (55)\\
VLA & 6.0 GHz & 330 (20) \\
\enddata
\tablecomments{All magnitudes are given in the AB system and are corrected for Galactic extinction.  Radio fluxes are in $\mu$Jy and the unabsorbed X-ray flux is in erg s$^{-1}$ cm$^{-2}$.}
\end{deluxetable}
\capstarttrue

\subsection{Radio and X-Ray Observations:  An Active Galactic Nucleus (AGN) Origin}

As described in \citet{DECamPaper5} and \citet{DECamPaper6}, we obtained radio and X-ray observations of \gw\ with the Very Large Array (VLA), Atacama Large Millimeter/submillimeter Array (ALMA), and \textit{Chandra}.  We detect host galaxy emission in all of these observations and list the X-ray and radio fluxes in Table \ref{tab:phot}.  For the X-ray emission we measure a best-fit power-law spectrum with $\Gamma = 1.2 \pm 0.2$, Galactic absorption with $N_H\approx 7.84 \times 10^{20}$ cm$^{-2}$ \citep{Kalberla2005}, and negligible host galaxy absorption, leading to an unabsorbed flux of $1.7\times 10^{-14}$ erg s$^{-1}$ cm$^{-2}$ ($2 - 10$ keV), which corresponds to a luminosity of $L_X\approx 3.2\times 10^{39}$ erg s$^{-1}$.  Assuming the X-ray emission is due to star formation activity, we find a star formation rate (SFR) of ${\rm SFR}\approx 1$ $M_{\Sun}$ yr$^{-1}$ using the SFR-$L_{X}$ relation from \citet{Grimm2003}, which is about two orders of magnitude higher than the well-determined value from the broadband SED modeling in \S\ref{sec:sed}.  

Similarly, we detect unresolved radio emission ($\lesssim 0.2''$, or $\lesssim 40$ pc) from the nucleus of \ngc\ with flux densities of about $330\pm20$ $\mu$Jy at 6 GHz and $210\pm20$ $\mu$Jy at 97.5 GHz.  The radio-millimeter spectral index is $\beta\approx -0.25$, shallower than observed in star forming galaxies, but consistent with AGNs.  In addition, the SFR estimated using $L_\nu (6\,{\rm GHz})\approx 6.8 \times 10^{26}$ erg s$^{-1}$ Hz$^{-1}$ is $\approx 0.1$ $M_{\Sun}$ yr$^{-1}$ \citep{YC2002}, again an order of magnitude in excess of the value from SED modeling. 

Thus, the radio and X-ray emission point to the presence of a low-luminosity AGN.  Using the stellar mass of \ngc, inferred from our SED modeling (\S\ref{sec:sed}), and the relation of \citet{RV2015}, we infer a supermassive black hole mass of $M_{\rm BH}\sim 10^{8.5}$ M$_{\Sun}$.  The X-ray luminosity therefore corresponds to $\sim 10^{-7}$ $L_{\rm Edd}$.  For this black hole mass, the ratio of the X-ray to radio luminosity is consistent with the fundamental plane of black hole activity \citep{Merloni2003}.  Furthermore, the morphology (\S\ref{sec:morph}) and age (\S\ref{sec:sed}) of \ngc\ are typical of low-luminosity AGN hosts \citep{Kauffmann2003}. 

\subsection{Optical Spectra}

During the course of obtaining optical spectra of the EM counterpart \citep{DECamPaper3} we also obtained spectra of the host galaxy.  Here, we use a spectrum obtained 1.5 days after the GW trigger with the Southern Astrophysical Research Telescope (SOAR) equipped with the Goodman High Throughput Spectrograph \citep{Goodman}.  Observations were carried out with the 400 l/mm grating and 1'' slit (R$\sim$830; see \citet{DECamPaper3} for details). We extracted the flux from \ngc\ in an aperture of width 32 pixels around the galaxy center, corresponding to the central $\sim$5''. The spectra were reduced using standard IRAF routines for bias and flat-field corrections, background subtraction, and wavelength calibration. Relative flux calibration was achieved using a standard star observation on the same night; our spectral analysis is not sensitive to the absolute flux calibration.       

\section{Morphological Properties of \ngc}
\label{sec:morph}

\capstartfalse
\begin{deluxetable}{lcc}
\tabletypesize{\small}
\tablecolumns{12}
\tablewidth{0pc}
\tablecaption{Measured and Derived Properties of \ngc\ and the Offset of the Optical Counterpart of \gw\
\label{tab:gal}}
\tablehead{
\colhead{Property}			 &
\colhead{Optical}		&	
\colhead{NIR}  }
\startdata
$n$ & 3.9 (0.4) & 5.1 (0.3) \\
$r_e$ (arcsec) & 16.2 (0.7) & 18.1 (1.6) \\
$r_e$ (kpc) & 3.3 (0.1) & 3.7 (0.3) \\
$\sigma_{\rm OT}$ (arcsec) & 0.0017 & 0.0006 \\
$\sigma_{\rm gal}$ (arcsec) & 0.0006 & 0.0001 \\
$\delta$R.A. (arcsec)& $-5.1796$ & $-5.1730$ \\
$\delta$Decl. (arcsec) & $-8.9208$ & $-8.9265$ \\
Offset (arcsec) & 10.315 (0.007) & 10.317 (0.005) \\
Offset (kpc) & 2.125 (0.001) & 2.125 (0.001) \\
Offset ($r_e$) & 0.64 (0.03) & 0.57 (0.05) \\
Fractional Flux & 0.54 & \nodata \\
\hline
\multicolumn{3}{c}{Derived Parameters} \\
\hline
$A_{V}$\tablenotemark{a} & \multicolumn{2}{c}{$<0.11$} \\
log(SFR$_{\rm 100Myr}$/$M_{\Sun}$ yr$^{-1}$) & \multicolumn{2}{c}{$-2.00^{+0.44}_{-0.66}$}  \\
log($M_{*}/M_{\Sun}$) & \multicolumn{2}{c}{$10.65^{+0.03}_{-0.03}$}  \\
log($M/M_{\Sun}$) & \multicolumn{2}{c}{$10.90^{+0.03}_{-0.03}$}  \\
$t_{\rm half}$ (Gyr) & \multicolumn{2}{c}{$11.2^{+0.7}_{-1.4}$}  \\
$t_{\rm age,spec}$ (Gyr) & \multicolumn{2}{c}{$13.2^{+0.5}_{-0.9}$}  \\
${\rm [Fe/H]}$ & \multicolumn{2}{c}{$0.08^{+0.02}_{-0.03}$}  \\
${\rm [Mg/Fe]}$ &  \multicolumn{2}{c}{$0.20^{+0.03}_{-0.02}$} \\
\enddata
\tablecomments{S\'{e}rsic parameters are from GALFIT.  Optical and NIR columns are averages of the values from the optical and NIR \textit{HST} observations in several filters.  ${\rm [Fe/H]}$, ${\rm [Mg/Fe]}$, and $t_{\rm age,spec}$ are from modeling of the spectrum and all other derived properties are from modeling of the SED.}
\tablenotetext{a}{95\% upper limit}
\end{deluxetable}
\capstarttrue

To determine the morphological properties of the host galaxy of \gw, we measured and fit the surface brightness profile using the \textit{HST} observations.  We used GALFIT v3.0.5 \citep{phi+10} to fit the 2D surface brightness profile of \ngc\ with a standard S\'{e}rsic function that is parameterized by the S\'{e}rsic index, $n$, the effective radius, $r_{e}$, and $\mu_{e}$, the surface brightness at $r_{e}$.  For comparison, we also used the {\tt ellipse} task in IRAF to fit isophotes to the galaxy images and then fit the resulting 1D isophotal intensity profile with a S\'{e}rsic function.  In both methods, the S\'{e}rsic model is defined in such a way that $r_{e}$ corresponds to the half-light radius.  The best-fit S\'{e}rsic parameters from GALFIT are listed in Table~\ref{tab:gal}.  

In general, we find that the surface brightness profile of \ngc\ can be well-described by a single $n \sim 3.9$ S\'{e}rsic component with $r_{e} \sim 3.3$ kpc and modest ellipticity (axis ratio $\sim 0.85$).  After subtracting the best-fitting GALFIT models from our data, the residual images suggest the presence of shell and asymmetric structure that are especially prominent in the F160W image (see Figure \ref{fig:hst}).  There are at least four concentric shells apparent in the F160W residual image with clear boundaries.  In Figure \ref{fig:hst}, we also show the residual image in the F475W filter, zoomed to show the complex dust structure surrounding the nucleus.  There is a large, approximately a few kiloparsecs, dust lane in a stretched out "s" shape, which appears to be connected to a smaller scale, $\sim$ 0.1 kpc, dust ring surrounding the brightest pixels.  Both the shell and dust structures may be indicative of past galaxy mergers \citep[e.g.][]{HQ1988,KD1989}.     

\section{Location of the Optical Counterpart}

\subsection{Offset}
\label{sec:offset}

To pinpoint the location of the optical counterpart relative to its host galaxy we measure its offset from the center of \ngc. For each {\it HST} image in which the optical counterpart is detected, we use {\tt SExtractor} to determine the uncertainty in the host galaxy center ($\sigma_{\rm gal}$) and the uncertainty in the optical counterpart location ($\sigma_{\rm OT}$), setting {\tt DEBLEND MINCONT} = 0.0005 to detect the optical counterpart against the high background emission from the galaxy. We then calculate for each filter the angular and physical offset, as well as the offset normalized by the effective radius, as determined from our surface brightness profile fitting (\S\ref{sec:morph}); see Table~\ref{tab:gal}.  We find a weighted mean offset of $2.125 \pm 0.001$ kpc, averaged over all filters.  For the normalized offset we find weighted mean values in the optical and NIR bands of $R/r_{e}=0.64\pm0.03$ and $0.57\pm0.05$, respectively, indicating that the optical counterpart is located within the half-light radius of the galaxy.  The smaller normalized offset in the NIR reflects the more extended surface brightness distribution at these wavelengths.

\subsection{Fractional Flux}

To determine the brightness of the galaxy at the location of the optical counterpart with respect to the overall host light distribution, we calculate the fraction of total galaxy light in pixels fainter than the optical counterpart position (``fractional flux''; \citealt{fls+06,FB2013,Blanchard2016}); this is a commonly measured quantity in the context of GRB host galaxies. We utilize the archival {\it HST}/F606W image to measure the galaxy brightness at the optical counterpart's location, which is localized to a single pixel, and create an intensity histogram for the entire host galaxy.  We consider pixels with a $1\sigma$ brightness level above the Gaussian sky brightness distribution to be part of the host galaxy light (e.g., \citealt{fls+06,FB2013,Blanchard2016}). We then determine the fraction of galaxy light in pixels fainter than the flux from the location of the optical counterpart.  In this manner, we calculate a fractional flux value of $0.54$. This value indicates that the galaxy brightness at the location of the optical counterpart is about average. We note that $\approx 75-80\%$ of short GRBs occur in fainter regions of their hosts than \gw\ \citep{DECamPaper8}.  

\section{Host Galaxy SED and Spectral Modeling}
\label{sec:sed}

We model the UV to mid-IR SED of \ngc\ using Prospector-$\alpha$, a 14-parameter galaxy SED model \citep{Leja2017a} that is built in the Prospector inference framework (B. Johnson et al. 2018, in preparation) and is optimized to fit UV$-$IR galaxy broadband photometry. Prospector-$\alpha$ uses a Bayesian Markov Chain Monte Carlo (MCMC) approach to modeling galaxy SEDs. In brief, the model fits a six-component non-parametric star formation history (SFH), a two-component dust attenuation model with a flexible attenuation curve, stellar metallicity, and a flexible dust emission model powered via energy balance.  Nebular line and continuum emission are added self-consistently through use of CLOUDY model grids from \citet{Byler2017}. This fit additionally includes a mid-IR AGN component described in \citet{Leja2017b}.  SFR and stellar mass measurements from this fitting assume a Chabrier initial mass function.

\begin{figure*}[ht!]
\begin{center}
\includegraphics[scale=0.27]{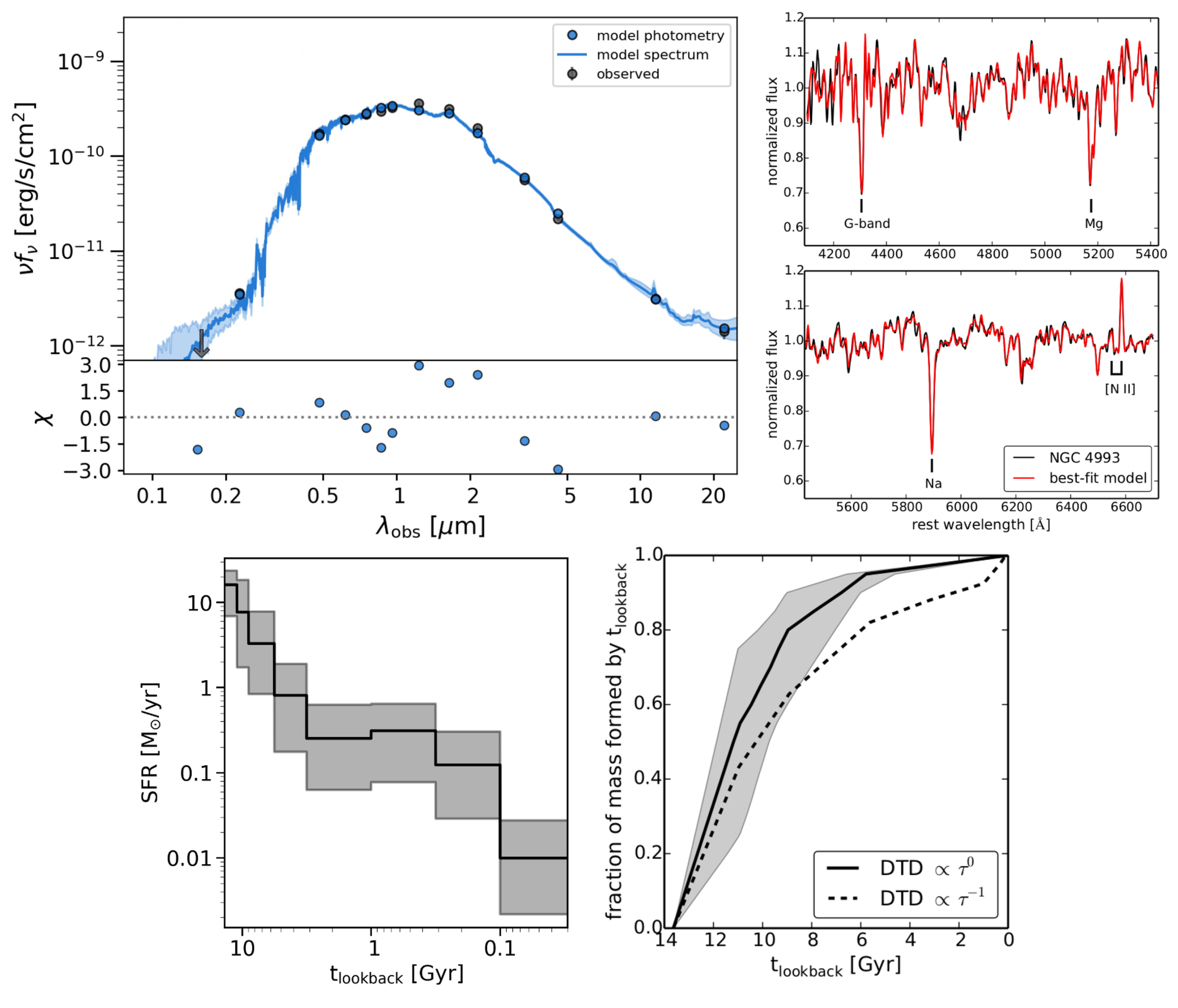}
\end{center}
\caption{\textit{Top Left}: Observed SED of \ngc\ (black circles) with the best-fit Prospector-$\alpha$ model (blue line; shaded region marks the 16th$-$84th percentile range).  \textit{Top Right}: Observed optical spectrum of the nucleus of \ngc\ (black line) with the best-fit spectral model (red line).  \textit{Bottom Left}:  The star formation history of \ngc\ from the best fit SED model (black line; shaded region marks the 16th$-$84th percentile range).  The SFH exhibits an overall exponential decline, with a very low present-day star formation rate.  \textit{Bottom Right}: Stellar mass build-up history (solid black line; shaded region marks the 16th$-$84th percentile range) as inferred from the SFH.  We find that $50\%$ of the stellar mass was formed by $11.2^{+0.7}_{-1.4}$ Gyr ago and $90\%$ was formed by $6.8^{+2.2}_{-0.8}$ Gyr ago.  Without prior knowledge of the intrinsic DTD of BNS mergers, the build-up history represents a proxy for the merger time probability distribution.  The dashed black line represents the resulting merger time probability distribution obtained by weighting the SFH with a $\tau^{-1}$ DTD truncated at 0.1 Gyr, which slightly shifts the distribution toward shorter merger times.  The uncertainty region is similar to that for the solid line.}
\label{SED}
\end{figure*}

In Figure \ref{SED}, we show the observed SED and best-fit model.  We find a low current star formation rate (averaged over the last 100 Myr) of log(SFR$_{\rm 100\,Myr}$/$M_{\Sun}$ yr$^{-1}$) $= -2.0^{+0.4}_{-0.7}$, a total mass formed in stars of log($M/M_{\Sun}$) $= 10.90^{+0.03}_{-0.03}$, and a stellar mass of log($M_{*}/M_{\Sun}$) $ = 10.65^{+0.03}_{-0.03}$ (defined as the current mass in stars and stellar remnants).  There is no significant dust extinction with a 95\% upper limit of $A_{V} < 0.11$.  Of particular interest here is the SFH, shown in eight temporal bins in Figure \ref{SED}.  We find an exponentially declining SFH with a peak star formation rate $\approx 10$ Gyr ago of about $10$ $M_{\Sun}$ yr$^{-1}$.  The SFH prior moderately favors a continuous star formation rate. The declining SFH in the posterior is thus driven by the photometry rather than the model priors. 

Using the SFH, we can calculate the fraction of stars produced by a given time to obtain the stellar mass build-up history, which we also show in Figure \ref{SED}.  Half of the stellar mass was formed by 11.2$^{+0.7}_{-1.4}$ Gyr ago ($t_{\rm half}$, the half-mass assembly time), due to the high SFR at early times, and 90\% was formed by 6.8$^{+2.2}_{-0.8}$ Gyr ago.  We list the main physical parameters resulting from the SED modeling in Table~\ref{tab:gal}.  

We also model the optical spectrum of \ngc\ with the {\tt alf} stellar population synthesis modeling code \citep{Conroy2012a,Conroy2017}, a two-component star formation history, the metallicity, and the abundances of 18 different elements.  This complex model space is fit with MCMC techniques, with the continuum shape removed with high-order polynomials.  For the present analysis we focus on three key quantities: the mass-weighted age, [Fe/H] metallicity, and [Mg/Fe], each of which is well-constrained by the data.  The data and best-fit model are shown in Figure~\ref{SED}; the model provides an excellent fit.  From the posterior distributions of the fitted parameters, we find a median mass-weighted age of 13.2$^{+0.5}_{-0.9}$ Gyr, a median metallicity of ${\rm [Fe/H]} = 0.08^{+0.02}_{-0.03}$, and ${\rm [Mg/Fe]} = 0.20^{+0.03}_{-0.02}$; the age inferred here is consistent with the SED modeling results.

\section{BNS Merger Timescale, Initial Separation, and Kick Velocity}

Using the SFH determined from the SED modeling, we can infer a probability distribution for the BNS merger timescale, and hence the initial binary separation.  We note that the inspiral timescale dominates over the stellar evolution timescale (which is at most tens of Myr).  The cumulative stellar mass build-up history, shown in Figure~\ref{SED}, can therefore be interpreted as the integral of the merger timescale  probability distribution.  The inferred old age of the stellar population and the exponentially declining SFH, lead to a median merger time of 11.2$^{+0.7}_{-1.4}$ Gyr and a 90\% confidence interval of $6.8 - 13.6$ Gyr.  Due to the lack of observational constraints on the intrinsic population delay time distribution (DTD), this interpretation makes the assumption that all merger times are equally probable.  To check the influence of this assumption on the result, we recalculate the merger time probability distribution with the SFH weighted by a $\tau^{-1}$ DTD (truncated at 0.1 Gyr), which results in a median merger time of $10.3^{+1.1}_{-0.8}$ Gyr (Figure~\ref{SED}).  While slightly shorter, the resulting merger time estimate is not significantly changed, due to the exponentially declining SFH.             

\begin{figure}[t!]
\begin{center}
\includegraphics[scale=0.4]{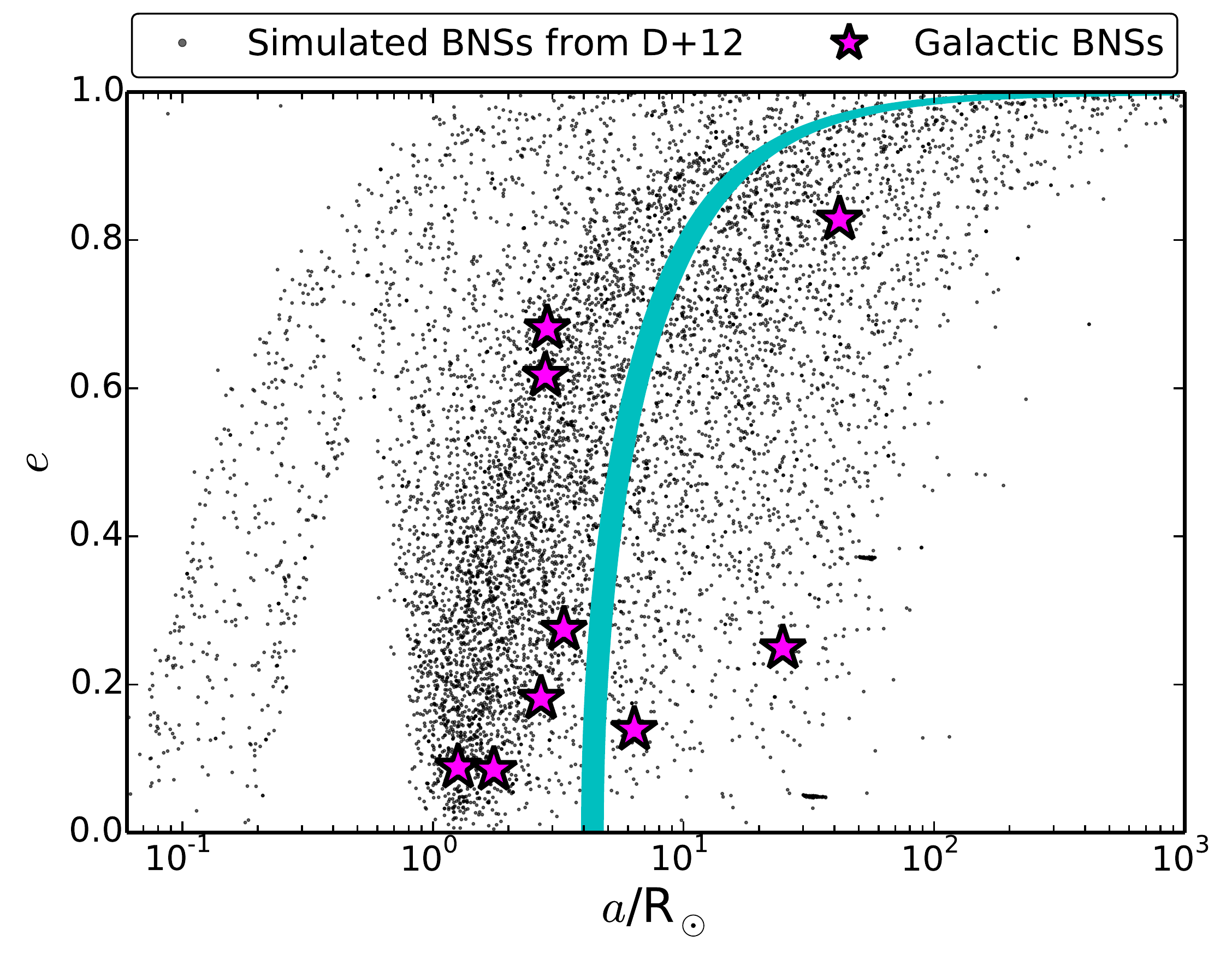}
\end{center}
\caption{Initial separation versus eccentricity plane with a shaded contour (cyan) representing the $90\%$ confidence interval for the merger timescale of the progenitor of \gw.  Magenta stars and black points represent the current $a$ and $e$ for Galactic BNS systems \citep[][and references therein]{VT2003,Wong2010} and a population of simulated BNS systems from \citet{Dominik2012}, respectively.}
\label{aeplot}
\end{figure}

The merger timescale depends on both the initial separation ($a_0$) and eccentricity ($e_0$) of the system \citep{Peters1964}:  
\begin{equation}
\label{t_merg}
\tau_{\rm merg}(a_0,e_0) = \frac{12}{19}\frac{c_0^4}{\beta}\int_{0}^{e_0}\frac{e^{29/19}[1+(121/304)e^2]^{1181/2299}}{(1-e^2)^{3/2}}de, 
\end{equation}
where 
\begin{equation}
c_0 = \frac{a_0(1-e_0^2)}{e_0^{12/19}}[1+(121/304)e_0^2]^{-870/2299}
\end{equation}
and $\beta$ is a constant related to the total and reduced mass.  Here we assume nominal neutron star masses of $1.4$ $M_{\Sun}$ since at the time of writing the measured neutron star masses from the GW data were not publicly available.  We can therefore convert our inferred median and 90\% confidence merger timescale to $a_0$ as a function of $e_0$; see Figure~\ref{aeplot}.  For $e_0\lesssim 0.5$ we find $a_0\approx 3.9-4.7$ R$_{\Sun}$ with a median of $\approx 4.5$ $R_{\Sun}$.  However, if the initial eccentricity of the system was large ($e_0\gtrsim 0.8$), then the initial separation could be tens of R$_{\Sun}$.       

In Figure~\ref{aeplot}, we also show the initial separations and eccentricities for the simulated population of BNS systems from \citet{Dominik2012} using their standard model (sub-model A) with solar metallicity, appropriate for \ngc.  The population has an overdensity roughly centered along the contour for $\sim 100$ Myr, though there is considerable spread in the initial separations and thus timescales for a given eccentricity.  At low eccentricity ($e_{0} \lesssim 0.2$) the distribution of  separations peaks at about a factor of $2-3$ less than the separation of the progenitor of \gw, though even at these eccentricities the spread is large.  Interestingly, the region of $a_0-e_0$ parameter space for the progenitor of \gw\ is mid-range compared to the Galactic BNS systems.

Combining the location of \gw\ within its host and the merger timescale probability distribution, we can assess the velocity imparted to the system due to a natal kick.  There are several unknown factors such as the formation location of the system, the fact that we can only measure a projected offset, and the exact gravitational potential of the host galaxy.   However, given the long merger timescale and the measured location within the host's half-light radius, it is most likely that any natal kick was not strong enough to unbind the binary from the host on an escaping trajectory.  Assuming the progenitor system was born near or within its current small offset, we can use the central stellar velocity dispersion ($\sigma_{*}$) to set an upper limit on the kick velocity.  The velocity dispersion is only marginally resolved in the host spectrum, yielding a nominal value of $\sigma_{*} \approx 150$ km s$^{-1}$ from the model fitting.  As the dispersion is less than the instrumental resolution, we consider the instrumental resolution to be a conservative upper limit on the velocity dispersion and therefore the kick velocity, which yields $v_{kick} \lesssim 200$ km s$^{-1}$.  This value is within the range of observed and simulated velocity distributions for Galactic BNS systems \citep{Wong2010,Tauris2017}.

Finally, we assess the possibility of a globular cluster origin for the progenitor of \gw\ (e.g., \citealt{lr10}).
We compare the limit on an underlying point source at the location of the progenitor ($\gtrsim -7.2$ mag) to the globular cluster luminosity function (GCLF) for giant elliptical galaxies, which peaks at $M_V\approx -7.4$ mag, with the brightest observed system at $M_V\approx -10$ mag \citep{hap+91}.  \citet{sbs+06} found that the luminosity at the peak of the GCLF remains the same regardless of galaxy size for elliptical galaxies.  Thus, the archival limit rules out the brighter half of the GCLF. Additional, deeper observations will be needed to definitively rule out a globular cluster at the position of the optical counterpart.

\section{Conclusions}

We presented new and archival data for \ngc, the host galaxy of the first BNS merger discovered through GW emission, and the first with an EM counterpart.  Using these data we investigated the location of the progenitor within its host, determined critical properties of the galaxy and its stellar population, and placed constraints on the merger timescale, initial separation, and kick velocity of the BNS system.  Our key findings are:
\medskip
\begin{itemize}
\item The host galaxy of \gw\ is an elliptical galaxy that is well-described by a $n\approx 3.9$ S\'{e}rsic profile in the optical, but with significant fine shell structure in the NIR indicative of past galaxy mergers.  We detect \ngc\ with X-ray and radio luminosities that suggest the presence of a weak AGN.

\item The offset of the BNS system from the nucleus of \ngc\ is $2.1$ kpc, with a normalized offset of 0.64 in the optical and 0.57 in the NIR, indicating that the merger took place within the host's half-light radius. The fractional flux value is $0.54$, consistent with this conclusion. 

\item From modeling the UV to MIR SED we find an exponentially declining SFH, with a median stellar population age of 11.2$^{+0.7}_{-1.4}$ Gyr.  The present-day SFR is low, $\approx 0.01$ M$_\odot$ yr$^{-1}$.

\item The median merger timescale is therefore 11.2$^{+0.7}_{-1.4}$ Gyr for the progenitor of \gw, with a 90\% probability the BNS system formed between $6.8-13.6$ Gyr ago.  Assuming a circular orbit and equal neutron star masses of 1.4 $M_{\Sun}$, this corresponds to an initial separation of $3.9-4.7$ $R_{\Sun}$ with a median of 4.5 $R_{\Sun}$; for large initial eccentricity the separation could be tens of $R_{\Sun}$.

\item Given the long merger timescale and small projected offset, we conclude that the binary system experienced at most a modest natal kick, with an upper limit of $200$ km s$^{-1}$.

\end{itemize}

This Letter demonstrates the utility of detailed host galaxy studies for inferring the properties of BNS systems.  Studying the host galaxies of future BNS mergers discovered by ALAV and localized by EM follow-up will allow for an observational measurement of the population DTD, a key parameter that informs the viability of BNS mergers as the dominant source of $r$-process enrichment in the Universe.  Comparison of the DTD as well as separations, natal kicks, and merger rates with population synthesis models will also yield great insight into uncertain phases, such as the common envelope phase, of massive star binary evolution.   

\acknowledgments
The Berger Time-Domain Group at Harvard is supported in part by the NSF through grants AST-1411763 and AST-1714498, and by NASA through grants NNX15AE50G and NNX16AC22G.  P.K.B. is grateful for support from the National Science Foundation Graduate Research Fellowship Program under grant No. DGE1144152.  D.A.B. is supported by NSF award PHY-1707954.  This research has made use of the NASA/IPAC Extragalactic Database (NED), which is operated by the Jet Propulsion Laboratory, California Institute of Technology, under contract with the National Aeronautics and Space Administration.  Some of the data presented in this Letter were obtained from the Mikulski Archive for Space Telescopes (MAST). STScI is operated by the Association of Universities for Research in Astronomy, Inc., under NASA contract NAS5-26555. Support for MAST for non-HST data is provided by the NASA Office of Space Science via grant NNX09AF08G and by other grants and contracts.

\facilities{HST (ACS, WFC3), SOAR, VLA, ALMA, CXO}

\bibliographystyle{yahapj}
\bibliography{BNS_host_paper}

\end{document}